\documentstyle[11pt]{article}
\def\ll{\label}
\def\re{\ref}
\def\c{\cite}

\def\r1{(\ref{$1})}

\def\th{\theta}
\def\ba{\begin{array}{c}}

\def\ea{\end{array}}
\def\pr{\prod}

\def\si{\sigma}

\def\De{\Delta}
\def\de{\delta}
\def\bet{\beta}
\def\ov{\over}
\def\ha{{1\over 2}}

\def\l{\left}
\def\l({\left(}
\def\r){\right)}
\def\r{\right}

\def\la{\lambda}
\def\al{\alpha}

\def\be{\begin{equation}}
\def\bc{\begin{center}}
\def\ec{\end{center}}
\def\bit{\begin{itemize}}
\def\eit{\end{itemize}}
\def\ee{\end{equation}}
\def\ed{\end{document}}
\def\bea{\begin{eqnarray}}
\def\eea{\end{eqnarray}}
\def\efr{\end{flushright}}

\begin{document}
\title{
Unifying quantization for  inhomogeneous integrable models}

\author{
Anjan Kundu \footnote {email: anjan.kundu@saha.ac.in} \\  
  Saha Institute of Nuclear Physics,  
 Theory Group \\
 1/AF Bidhan Nagar, Calcutta 700 064, India.
\\ Phone: +91-33-23375346(-ext. 2365)\\
Fax: +91-33-23374637  }
\maketitle
\vskip 1 cm

\begin{abstract} 

 Integrable inhomogeneous versions  of the models like
NLS, Toda chain, Ablowitz-Ladik model etc., though  
  well known at the  classical level, have never been investigated for
 their possible quantum extensions.
  We propose a unifying scheme for constructing and solving such quantum
integrable
   inhomogeneous  models including a novel
 inhomogeneous sine-Gordon model, which  avoid  the difficulty related to
  the customary non-isospectral flow by introducing the inhomogeneities
 through some central elements of the underlying algebra. 
\end{abstract}

\smallskip

  
\noindent{\large {\bf 1. Introduction}}

Over more than  past two decades  active interest has been focused in the
study of inhomogeneous integrable models. Various types of space and time
inhomogeneities have been introduced successfully in the well known
classical integrable models like  
 the nonlinear Schr\"odinger equation (NLS)
through linear, quadratic, cylindrical,  radial etc functions
\c{chlu,cade,ba,rala2}, the Toda chain (TC) with  arbitrary
inhomogeneity  \c{rale} and the
 Ablowitz-Ladik (AL) model  with linear in space and  arbitrary in time
functions
\c{br3,bi2,ko2} etc., preserving  their
integrability. The associated Lax operators, soliton solutions, Painlev\'e
integrability criteria etc. have been extensively studied for such systems
\c{qule,porsez97}.  Such inhomogeneities might induce intriguing effects
 like locally varying interactions, space-time dependence of
 soliton velocities, trapping of solitons in periodic movements  etc. In the
simplest example of linearly inhomogeneous NLS
   the solitons  move with a uniform acceleration \c{chlu}.

 However,   it is rather surprising that,  investigations
 for the above inhomogeneous models have been carried out only 
at the classical level and no systematic effort
has been made
 toward their quantization, except perhaps our  
  own preliminary study \c{kuninhom99}. A possible reason for this 
 might be,  that 
   the inhomogeneities in such models are believed to be  generated through 
non-isospectral flows with  space-time dependent  spectral
parameter $ \la=\la (x,t)$. This in general
   would lead   to a dynamical classical $r$-matrix and
  consequently to a quantum $R$-matrix with space-time dependence,
 taking us  beyond the known
   formulation of quantum integrable systems.

 We   overcome  this difficulty  for some of the nonisospectral
flows and  propose  a unifying 
 quantization scheme for the integrable inhomogeneous NLS, TC and  AL
 models together with  a novel  inhomogeneous sine-Gordon model.
  The idea of such quantization is based on our earlier scheme
\c{kun99} as well as  on the observation that,
   certain  non-isospectral  problems may be looked from a bit different
 angle by
     considering  spectral parameter $\la$ still to be a constant, while
  relegating
 the  inhomogeneities  to a set of central elements of an
underlying algebra, which  ensures   quantum integrability.
 These inhomogeneity elements, which  may be interpreted   
as  external classical fields, 
commute with all other basic operators and  appear  only 
in the quantum   Lax operators, but    not in the quantum $R$-matrix.
Such fields might have   arbitrary
 space-time dependence with suitable restrictions
 on their boundary and   asymptotic conditions, imposed by the
integrability.
 A positive outcome of this approach is that,
the classical and the quantum $R$ matrices for such
   integrable inhomogeneous models remain the same as in
  their homogeneous counterparts.

 Note that the algebraic Bethe ansatz
 solution of quantum integrable systems 
 depends on the vacuum Lax operator as well as on the
$R$-matrix elements.  Therefore, since in our inhomogeneous
extensions,  the $R$-matrices are kept the same,
 the related result are affected only  by the 
inhomogeneity in their  Lax operators, which   signifies 
  the presence of impurities, defects,  density fluctuations in the
 media, or the influence of variable external fields, depending on the
 physical situations \c{qinhom}.

  We concentrate  here  on  inhomogeneous quantum NLS, 
  and TC models as well as   AL and SG
  models, in their  different forms. Note that homogeneous
versions of  all  these models are  
quantizable \c{fadrev,kulskly},
where the first two sets of
 models are 
linked to   the rational, while  the last two to 
 the  trigonometric  $R$-matrix.

The arrangement of this paper is as follows. We briefly review in sect. 2 
various known forms  of the classical inhomogeneous equations and in sect. 3 
the basic structures of the quantum integrable systems. We present in 
sect. 4  the construction  of  quantum integrable extensions of 
inhomogeneous models 
 in a unifying
way and  in  sect. 5 show their   exact and systematic solution
 through the   Bethe
ansatz   method. 
      Sect. 6 gives the concluding remarks.
 

\noindent{\large {\bf 2. Classical integrable  inhomogeneous models}

We  briefly list   different well  known  forms of the 
classically integrable 
inhomogeneous NLS, TC and AL models, where the details can be found in the
cited references.

\noindent {\bf 2.1 Inhomogeneous NLS equations}

\noindent 1.{\it NLS equation with linear inhomogeneity } (XNLS)

An integrable  NLS equation (NLSE) with 
 linear $x$-dependent inhomogeneity was 
proposed in \c{chlu,cade} 
with  non-isospectral flow  
$\dot {\Lambda }= \alpha$, $\Lambda (t) $ being the spectral parameter.
The inhomogeneous equation allows uniformly accelerated soliton solution as well as    time
dependent  wave-number and   frequency of the enveloping  wave
\c{chlu}.  

\noindent 2. {\it Cylindrical NLSE } (CNLS)

An integrable  cylindrically symmetric  NLSE having an explicit
time-dependent coefficient was proposed at the classical level, with
non-isospectral dependence of 
 spectral parameter:
$\Lambda (x,t)= {\lambda_0 \over t} +{x \over 4t} \ $ \c{rala2}.
%

\noindent 3. {\it NLSE  with t-dependent coupling }(TNLS)

 It is shown through Painlev\'e analysis 
  that the time-dependence of the nonlinear 
coupling in the integrable NLSE can be of
 the form $F(t)=  {1 \over at+b}  $.
This model  also has the same non-isospectral flow as the previous case.

\noindent 4. {\it Radially symmetric NLSE} (RNLS)  
and \\ 
5. $x$-{\it dependent  nonlocal  NLSE} (NLNLS)
are known   to be classically integrable \c{cade,rala2}. 

In both these cases the non-isospectral condition was taken to be 
$\dot \Lambda =a \Lambda ^2$ .

\noindent 6. { \it NLSE with more general inhomogeneity   } (FNLS)

 Integrable  inhomogeneous classical NLSE  
  with  more general  $x$-dependent coefficient 
 was proposed as  
$ i\psi_t+\psi_{xx}+2(|\psi|^2-F(x))\psi=0,
  $
where $F(x)$ can be linear, quadratic or in  more general form,
depending on the type of non-isospectrality given by  
 the space-time dependent 
spectral parameter
satisfying    $\Lambda_t=2
(\Lambda^2)_x F_x $  \c{ba}. 

\noindent{\bf 2.2 Inhomogeneous Ablowitz-Ladik model}

The AL model    was discovered first as a discrete
NLSE.  The inhomogeneous  AL model  
 was studied at the classical level in
a series of papers \c{br3,bi2,ko2},
which describes the system to be in an external  time-dependent 
and  linear in space potential and
 induces  an intriguing effect of trapping the  soliton
and forcing it to a 
periodic movement. 

\noindent{\bf 2.3 Inhomogeneous Toda chain}

 Classically   integrable
inhomogeneous Toda chain with varied   space-time dependence,
 in of the form  \c{rale} 
$      u_{tt}(n)= g_1(n) e^{u(n-1)-u(n)}-g_1(n+1)e^{u(n)-u(n+1)}+
\dot g_2(n)+$
 {\t boundary terms},  
allows  different choices for inhomogeneity  functions
 $g_a(n,t), a=1,2 $.



\noindent {\large {\bf 3. Quantum integrable systems}}

 As is well known \c{fadrev},
the quantum integrability of a  system is guaranteed  
by the  quantum Yang-Baxter equation (YBE) 
\be {R}(\la-\mu)L_{j}(
\lambda) \otimes L_{j}( \mu) = (I\otimes L_{j}( \mu))(L_{j}( \lambda)\otimes
I) {R}(\la-\mu)
,\ll{qybel} \ee
with the
  quantum  Lax operator $L_j$ 
of  the discretized   model  defined at all sites $j=1,2,\ldots,N$ and 
 a site-independent   $R$-matrix, together
with the  ultralocality condition: $
 (I\otimes L_{k}( \mu))(L_{j}( \lambda)\otimes
I)=L_{j}(
\lambda) \otimes L_{k}( \mu), \ j \neq k $.
The  corresponding classical model with the related  $r$-matrix: $R(\la) \to
I+\hbar r(\la)+O(\hbar ^2) $
 would satisfy  the classical  YBE
$ \{L_j(\lambda)\otimes _, L_k( \mu)\}= \de_{jk}
[{r}(\la-\mu),L_j( \lambda)\otimes  L_j( \mu)] 
$
for the discretized Lax operator $L_j( \lambda) $, which 
  goes  to  the corresponding
field Lax operator $U(x, \lambda) $  at the continuum limit $\De \to
0$: 
$L_j( \lambda) \to I+i \De U( x, \lambda) $. In such an  
 approach therefore only the space-Lax operator  $L_j(\la)$ or
its continuum version $U(x, \la)$ together with the canonical relation
 play the central
role, while the time-Lax operator  $V(x,\la) $ becomes insignificant. Moreover
in quantum problems, unlike  
  classical equations of motion,  the main emphasis
 is on solving the eigenvalue problem of the Hamiltonian: $H|n> =E_n|n> , $
together  with that of other  conserved operators $C_j, j=1,2,\ldots$.
These conserved operators are generated by the transfer matrix 
$\tau(\la)=tr (\prod_j L_j (\la)) $  through expansions    $
\tau(\la)=\sum C_{\pm j} \la^{\pm j}$ or through similar expansions of
  $\log \tau(\la)$. 

The algebraic Bethe ansatz, an exact method for quantum integrable systems
solves therefore a  general eigenvalue problem: $\tau(\la)|n>
=\Lambda_n|n> $.
 As an essential condition for the  quantum integrability
, which follows  from
the   ultralocality and the quantum YBE (\re{qybel}) 
the conserved operators including the Hamiltonian must commute 
   mutually
  \c{fadrev}.

It is important to note that, the YBE (\re{qybel}) not only
 ensures the  quantum integrability but also   plays  a
 central role in the exact Bethe ansatz (BA) solution and
for this it is essential that, 
  the  $R$-matrix appearing in it must  not depend on the 
space-time coordinates. Consequently, since the 
 $R(\la-\mu)$-matrix is a function of  $\la-\mu$,
  the spectral parameter 
 can not  dependent  on space-time variables, even in 
inhomogeneous models. 
 Therefore,   for quantization of our inhomogeneous models we introduce
inhomogeneity  through  Lax operator $L_j$,
  by keeping the spectral parameter
constant and demanding that the changed
 Lax operator must satisfy the YBE (\re{qybel})  with
  the same ${R}$-matrix  
  as in the original homogeneous model. A similar argument holds 
for the    classical YBE
at the classical limit. Therefore,
it should be clear that we can consider  inhomogeneous  extensions 
for only those  models, which are quantum integrable at their homogeneous
limit.  The   inhomogeneous  versions of
 integrable NLSE, TC and AL models discussed above  are known
    only at the 
classical level  as a  non-isospectral problem, whereas 
 inhomogeneous  SG model seems to not have been studied even at the
classical level. 
 Our main concern now    would be to look into  the
 possible  quantum  extensions of these inhomogeneous models, which
apparently have
never been undertaken. 

\noindent {\large {\bf 4. Unifying quantization  for inhomogeneous models}}

We formulate a quantization scheme for  inhomogeneous integrable  models
of both rational and trigonometric type based on an unifying quantum
algebra \c{kun99}. The inhomogeneities are introduced in these models
through central elements of the underlying algebra in a systematic way.\\
{\bf 4.1 Rational class of inhomogeneous models}:

Recall that  both the   homogeneous  NLS and  TC 
models are  quantum  integrable and
 associated with the well known rational $R(\la)$-matrix   \c{kulskly}
given by its nontrivial elements as 
\be a(\la) \equiv R^{11}_{11}=R^{22}_{22}= \la+\eta, \ 
b(\la) \equiv R^{12}_{12}
=R^{21}_{21}=\la, \ c \equiv
R^{12}_{21}
=R^{21}_{12}=\eta , 
 \ll{Rrat} \ee 
 For  constructing the 
  intended inhomogeneous extensions of these models we   start   from    
 a    rational discrete general Lax operator  \c{kun99}
 \be
L_{rn}{(\la)} = \left( \begin{array}{c}
 {c_1^0} (\la + {s_n^3})+ {c_1^1} \ \ \quad 
  s^-_n   \\
    \quad  
s^+_n    \quad \ \ 
c_2^0 (\la - {s^3_n})- {c_2^1}
          \end{array}   \right), \ll{LK} \ee
 which was shown to yield 
all  integrable models
 with $2 \times 2$ Lax operators,   belonging to the rational class.
Note that   $L$-operator (\re{LK}) 
depends on  
 operators ${\bf s}$ 
with generalized spin algebra 
\be  [ s^+_n , s^-_m ]
= \de_{nm}( 2m^+ s^3_n +m^-),\ \ \ \ 
  ~ [s^3_n, s^\pm_m]  =\pm  \de_{nm}( s_n^\pm) 
  \ll{k-alg} \ee
and a set of commuting operators $c^\al_a, \al=0,1, a=1,2$,  forming
central elements $m^+=c_1^0c_2^0,\  m^-= c_1^1c_2^0+c_1^0c_2^1$, which 
 commute with all other operators. 
We note that,   algebra (\re{k-alg}) 
  dictated by  YBE (\re{qybel}) 
guarantees the quantum integrability of all   models realized through the
Lax operator (\re{LK})  and 
 since the algebraic relations   are valid locally at all sites $n=1,2, \ldots
, $
 the central elements can  be taken in general to be time as well as 
 lattice-site 
 dependent   variables:  $ c^\al_{a,n}(t)$.
We would see in the sequel  that these  
 inhomogeneous  elements would  
in fact be crucial
in introducing inhomogeneity in integrable 
quantum systems we intend to construct.

\noindent  {\bf I. Quantum inhomogeneous NLS model}

It is easy to see that when  inhomogeneous elements are absent  
 i.e.  when $ c^0_1=c^0_2= c^1_1=-c^1_2=1$, giving $m^+=1, m^-= 0$,
 the  generalized spin algebra  (\re{k-alg}) reduces simply to 
$su(2)$ spin algebra and through
 bosonic realization of its generators  by the Holstein-Primakov
representation (HPR) we can recover from (\ref{LK})
 the Lax operator $L^{lnls}_n$ of  the exact
lattice version of the standard  quantum NLS,
 introduced by Korepin and Izergin
\cite{izko}.  
Therefore  for constructing the corresponding 
inhomogeneous quantum  model 
 our strategy   would be  to follow a  similar path, but with     
nontrivial    elements
 $ c^\al_{a,n}(t)$,   for which  we  first  find   
 a realization of 
(\re{k-alg})  in the form of generalized
 HPR 
 \bea
 & & s^3_n=s_n-N_n {\De }, \ \ \    s^+_n= f(N_n) \psi_n \sqrt {\De }, \ \ 
  s^-_n= \psi_n ^\dag f(N_n)\sqrt {\De }
, \nonumber  \\ \mbox{with} & & 
 f^2(N_n)=m^-_n+m^+_n (2s_n -N_n  {\De }), \ \ N_n=\psi_n^\dag
\psi_n.
\ll{ilnls} \eea
 where $ [\psi_n , \psi_m^\dagger ] =  {1 \ov  {\De }} \de_{nm}$,
 $s_n$  is the spin parameter taken also to be site-dependent
  and $\Delta $  is the  lattice spacing. 
 (\re{ilnls}) clearly recovers
 the standard HPR at the homogeneous limit. Therefore  in analogy with
\cite {izko} we  show that, the generalized HPR  
yields an exact lattice version
of the inhomogeneous quantum NLS model, for
which  we  demonstrate   first 
 that,  the associated $L$-operator (\ref{LK})   has a consistent 
 continuum limit, which 
 recovers
the  Lax operator of the corresponding 
quantum   field model  at  $\Delta \to 0$. 
Making a particular reduction
 of the central elements 
 as
\be c^0_1=c^0_2 \equiv g_n(t) , \ 
c^1_1=-c^1_2=f_n(t), \ \ \mbox{ giving} 
\ m^+=g_n^2, m^-= 0, \ee
 where $f_n, g_n$ are  space-time dependent
 arbitrary functions, we find    that,
  at the high  spin limit
 $s_n \to {1 \over \Delta} g_n^{-1}$  
  the generalized HPR (\re{ilnls})  reduces  
 (\re{LK}) 
to \be
 L^{lnls (inh)}_n(\la )=\De \sigma^3L_n(\la ) = I+\De \left( \begin{array}{c}
 \Lambda_n- g_n\De N_n  \ \ \quad 
  \psi_nf^{(0)}(N_n)\sqrt {g_n }  \\
-\psi^\dag_nf^{(0)}(N_n)\sqrt {g_n }   \quad \ \ 
-(\Lambda_n+ g_n\De N_n) 
          \end{array}   \right) \ll{Linls} \ee
where $\Lambda_n =g_n \la+f_n $ and $f^{(0)}(N_n)=( {2-g_n\De ^2 N_n
})^{\ha} $.
One can verify that the related discrete  model 
is a quantum integrable system, since   the   
 Lax operator (\re{Linls}) associated with it 
  together
with the rational $R$-matrix (\re{Rrat})    exactly 
 satisfy the quantum YBE
(\re{qybel}).
We  observe  further that, at the field limit  $\De \to 0$, when 
 $\psi_n(t) \to \psi (x,t), \ f_n \to f(x,t),\ g_n
\to g(x,t)$, the lattice Lax operator (\re{Linls}) reduces to:
  $L^{lnls (inh)}_n= I-i \Delta {U}^{nls (inh)}(x,\la) +O(\De ^2) $, 
  recovering the corresponding field Lax operator
\be
U^{nls (inh)}(x, \la ) =  \left( \begin{array}{c}
 i\Lambda   \ \ \quad 
  i\sqrt{2g}\psi  \\
-i\sqrt{2g}\psi^\dag  \quad \ \ 
-i\Lambda 
          \end{array}   \right) \ll{qU} \ee
  where $\Lambda=g(x,t) \la+f(x,t)$. 
We may check  again that,   (\re{qU})
 with  quantum field operators
 $[\psi(x,t), \psi ^\dag (y,t)]=\de (x-y) $ and  $ f(x,t),\ 
 g(x,t)$ acting  as classical background fields satisfies  
YBE  (\re{qybel}) with the same $R$-matrix (\re{Rrat}), up to the
first order in $O(\De )$, which however is sufficient for
the integrability of a quantum field
model. Therefore the quantum Lax operator
(\re{qU}) clearly represents  an   inhomogeneous  
  generalization of   the quantum integrable   
 NLS   field model.

For 
the  homogeneous lattice NLS model  it has been established   \cite {izko}
that,   at certain  values of the  spectral parameter  $\la =\nu_i $,
 where  the  quantum determinant ${\rm qdet} (L(\la))\equiv \ha tr
 (L_n(\la)\si ^2 L_n(\la+1)\si ^2) $  vanishes
   the   Lax operator is  
 realizable  as a projector as well as an inverse projector. This important
property was shown to be useful  for proving the local character of the  
 discrete
Hamiltonian, i.e.   its interaction spreading over to  only
few nearest neighbors.
For our 
  inhomogeneous 
  lattice NLS model  we could show that 
${\rm qdet} (L(\la))$ for the associated 
 Lax operator 
(\re{Linls}) vanishes at 
 $\nu_1=-{1\over \De} , \ \nu_2^{(n)}=(g_n \De -\nu_1) $ 
and the $L_n$-operator 
     at these points can   be represented indeed  as projectors and inverse
projectors as
 \be
L^{lnls (inh)}_{ij}(\nu_1 ) = \al_j\beta_i
\ \ \mbox{and} \ \ L^{lnls (inh)}_{ij}(\nu_2^{(n)}) = \tilde \al_j \tilde
\beta_i
 \ll{iLproj} \ee
 {with} $ 
\beta_1= \tilde \al_2=   \De  \psi_n\sqrt {g_n },
\ \ \ \al_1= \tilde \beta_2 =  - \De  \psi^\dag _n\sqrt {g_n } , \ \
 \al_2=  \beta_2= f^{(0)}(N_n), \ \ \  \tilde \al_1= \tilde  \beta_1=
f^{(0)}(N_n-1) $.

However we note that the point $\nu_2^{(n)} $
becomes site-dependent for this 
inhomogeneous  model and therefore the projector-representation can not be
achieved for all $L_n $ at a single value of $\la $, which would lead
inevitably to nonlocal interactions for its Hamiltonian in both 
classical as well as quantum cases. By choosing some particular forms for
$g_n$, e.g., taking them same for all odd $n=2k+1$, we can  partially
regulate the
localization of the Hamiltonian, though such inhomogeneous models in general
would lead to nonlocal interactions.

A further technique of pairwise grouping of the Lax operators at different
sites for achieving  locality for the Hamiltonian has been shown for the
homogeneous quantum lattice NLS. Whether the same  technique is applicable 
also to our corresponding inhomogeneous case needs more detailed analysis ,
which we leave  for future study.

At the classical limit, when  field operators become classical
 functions one can show that, 
(\re{Linls}) satisfies  classical YBE 
with the classical rational $r(\la)={1 \ov 2 \la }(I+\vec \sigma \cdot 
\vec \sigma ) $-matrix and 
 represents  therefore the   Lax operator  of a new
  integrable   discretization of    inhomogeneous classical NLSE.
Similarly,  at the  classical limit   (\re{qU}) must also
satisfy  the continuum version  of the classical YBE:
$ \ \{U(x, \la )\otimes _, U(y, \mu)\}= \de (x-y)
[{r}(\la-\mu),U(x, \la )\otimes I+I \otimes  U(x, \mu)], $
\  with the same $r$-matrix and  
would correspond to the 
classically integrable  inhomogeneous NLS field models.
 Notice that, if  in (\re{qU})  one    considers $\Lambda$ to be 
 the spectral parameter instead of $\la$  entering    $r(\la)$-matrix, the problem would look like a customary
non-isospectral flow with nontrivial $\Lambda_t, \Lambda_x$.  
In fact if we rename 
$\sqrt{2g}\psi =Q$, (\re{qU}) 
  would coincide  at its  classical limit with 
 the Lax operator of  known   inhomogeneous NLS models and
 hence would recover for
 different choices of the  
functions $f(x,t), g(x,t)$, the  inhomogeneous NLS equations
proposed earlier. 
For example, i) $g=1, f=\al t$ would give  linear
time-dependence of  $\Lambda=\lambda+\al t$ with $\dot \Lambda=\al$
reproducing the XNLS, 
 while ii) $g={1 \over t}, \ f= {x \over 4t}$  would give
the inhomogeneous CNLS. 
We can get also  the inhomogeneous TNLS
by  multiplying  the field variable in 
CNLS  by   a function of $t$, 
 which  however is  not a canonical step.
We may consider now a more general situation  iii) g=X(x), \ f=T(t)X(x),
which gives $\Lambda=X(x)(\la +T(t))=g(x)h(t)$, coinciding with the
separable solution of the  FNLS
 \c{ba}. Note 
that, the related non-isospectral picture
 satisfying the equation $ \dot T=X^{'} (T-\la)^2 +F^{'}(x)/X$, is
compatible with our isospectral relations.
 Therefore a nontrivial solution \c{ba} : $ X=a x+b, \dot T=a(T-\la)^2 +a_0,$
yielding the NLS with quadratic x-dependence: $ F(x)=a_0(\ha ax^2+bx)+ c$,
should also be reachable from our construction.

In cases of radial  and nonlocal NLS  however it
appears that,  due to nonlinear nature of the  non-isospectrality
 and more involved  noncanonical fields, the known classical models
RNLS and NLNLS 
would not coincide with   the classical limits of our quantum inhomogeneous
models. It is important to note that, in classical Lax pair approach
non-isospectrality with space-time dependence
 appears in both the space ($U$) and time ($V$) 
Lax-operators  and moreover   the allowed  transformations 
  may not respect  the  canonicity of the fields.
In the  quantum case  on the other hand the inhomogeneity 
can  appear only in the space part $U$ or 
$L_j$, since the analog of  $V$-operator is absent here. At the same time, 
  the canonical structure  defined  in quantum models is fixed 
by the associated    $R$-matrix and should  remain
unchanged under all transformations.
Nevertheless, as we have seen, apart from a simple noncanonical 
transformation of the field:
 $\psi \to Q=\sqrt{2g}\psi ,$ with $g$  a
space-time dependent  function, (\re{qU}) in the classical limit 
 maps into the
known  Lax operator 
for the inhomogeneous NLS equations.

\noindent {\bf II. Quantum inhomogeneous Toda chain}

We intend to construct the  quantum extension of the inhomogeneous Toda
chain starting again from the same Lax operator (\ref{LK}), but choosing now
the
 inhomogeneity fields $ c^1_{1j}, c^0_{1j}$ arbitrary, while 
$ c^1_{2}=c^0_{2}=0$. This clearly   gives  both $m_j^\pm=0$ 
reducing  algebra (\re{k-alg}) simply to
\be  [ s^+_n , s^-_m ]
= 0,\ \ \ \ 
  ~ [s^3_n, s^\pm_m]  =\pm  \de_{nm} s_n^\pm 
  \ll{algtc} \ee
which can be realized through canonical quantum fields
 $[u_n,p_m]=i\de_{nm} $ of the Toda chain  as  
\be s_j^3
= -i p_j ,\ \  s_j^\pm = (G_j)^{\pm 1}e^{\mp u_j},  \ll{k-b}\ee
with arbitrary function $G_j$. This 
  reduces   (\re{LK})  to   the quantum Lax operator
of the inhomogeneous  Toda chain:
 $L^{TC(inh)}_{j}{(\la)}=L_{j}{(\la)}/c^0_{1j} $ as
 \be
L^{TC(inh)}_{j}{(\la)} = \left( \begin{array}{c}
 \la  -i (p_j - g_{2(j)}) \ \ \quad 
  {e^{ u_j} F_{1(j)}}   \\
  e^{- u_j}F_{2(j)}   \quad  
    \quad \ \ 
0
          \end{array}   \right), \ll{qitc} \ee
with the inhomogeneous arbitrary  functions \be g_{2(j)}
=i({c^1_{1j}\over c^0_{1j}}), \ F_{1(j)} ={ 1/(c^0_{1j}G_j)}, 
, \ F_{2(j)} ={ G_j/c^0_{1j}}.  \ll{gff} \ee 
 These   space-time dependent
  functions  obviously  enter the  Hamiltonian of the inhomogeneous
quantum Toda chain  as
\be  
H=\sum_n^N P_n^2-g_{1(n+1)} e^{u(n)-u(n+1)}, \ P_n=p_n- g_{2(n)}, \
g_{1(n)}= F_{1(n-1)} F_{2(n)}
\ll{htc}\ee
and the conserved total momentum operator as $P=\sum_n P_n$, generated
through expansion of the transfer matrix $\tau(\la)$  using    
  the Lax   operator  (\re{qitc}) as $P=C_{N-1}, \ H=C_{N-2} $ etc.

At the classical limit one can derive the evolution equation
from  this inhomogeneous Hamiltonian (\re{htc}), which considering the
relation obtained here:
$-i p_j= \dot u_j- g_{2(j)}$ would 
recover the   known
 inhomogeneous classical  Toda chain  equations
for different choices of
 functions $g_{a(n)}, a=1,2$ \c{rale}. 

\noindent {\bf 4.2  Trigonometric  class of inhomogeneous models}:

since  both   AL and SG 
 models belong to the trigonometric  class, we focus 
 now on the  
idea behind the construction of  quantum integrable models 
associated with the well known trigonometric $R_{trig}$-matrix \c{kulskly,fad}, which is a
$q $-deformation of (\re{Rrat}) and given through its nontrivial elements as 
\be a(\la) \equiv R^{11}_{11}=R^{22}_{22}= \sin (\la+\eta), \ 
b(\la) \equiv R^{12}_{12}
=R^{21}_{21}=\sin \la, \ c \equiv
R^{12}_{21}
=R^{21}_{12}=\sin \eta , 
 \ll{Rtrig} \ee 
Following our earlier  work \c{kun99} 
we start from a  trigonometric generalization 
of (\re{LK}):
\be
L_t{(\xi)} = \left( \begin{array}{c}
  \xi{c_1^+} e^{i \al S^3}+ \xi^{-1}{c_1^-}  e^{-i \al S^3}\qquad \ \ 
2 \sin \al  S^-   \\
    \quad  
2 \sin \al  S^+    \qquad \ \  \xi{c_2^+}e^{-i \al S^3}+ 
\xi^{-1}{c_2^-}e^{i \al S^3}
          \end{array}   \right), \quad
          \xi=e^{i \alpha \la}. \ll{Lt} \ee
with the $q$-spin operators ${\bf S}$ satisfying 
a generalized  $q=e^{i \al}$-deformed algebra 
\be
 q^{S^3}S^{\pm} = q^{\pm 1} S^{\pm}q^{S^3} , \
 \ \ [ S^ {+}, S^{-} ] =  {- 
h} 
 \left ( c^+_1c^-_2 q^{2 S^3}-c^-_1c^+_2 q^{-2 S^3} 
 \right), \ \ h= {q -q^{-1}},
\ll{qa}\ee
with $c^\pm_a a=1,2 .$ 
One can see  that algebra (\re{qa}) is a generalization of the well
known quantum algebra $sl_q(2)$, which is   recovered easily 
 at the homogeneous limit 
$c^{\pm}_1=c^{\pm}_2=1$. 
It is interesting to show that the $L $-operator (\re{Lt}) together with
the $R_{trig}$-matrix (\re{Rtrig}) satisfy the quantum YBE (\re {qybel}), which 
yields algebra (\re{qa}) as  the condition for the  quantum 
integrability. It is also intriguing to note that,
 at the undeformed limit $ q \to 1, $ the trigonometric $L$-operator (\re{Lt})
goes to its rational limit (\re{LK}), while the quantum algebra (\re{qa})
reduces to the generalized spin  algebra (\re{k-alg}).
    The 
 construction of inhomogeneous models for this trigonometric class follows also 
the idea adopted above,
i.e. we consider all central elements $c^\pm_a, a=1,2$, appearing in 
(\re {Lt}), (\re {qa})  
to be space-time dependent functions, which   
may vary arbitrarily   at  different
lattice points $j$ and this leads 
  naturally 
 to  integrable  lattice models with inhomogeneity . However since the 
underlying algebra (\re{qa}) 
 does not change even 
 with such inhomogeneous extension,
  the system remains linked   to the same trigonometric 
$R$-matrix and hence  retains its exact solvability through the Bethe
ansatz.

\noindent {\bf III. Q uantum  inhomogeneous sine-Gordon model}

Note that an interesting 
 realization  of the  quantum algebra (\re{qa}) in the canonical
variables: $ [u,p]=i, $ 
may be given  as
\be
 S^3=u, \ \ \   S^+=  e^{-i p}g(u),\ \ \ \ \ 
 S^-=g(u) e^{i p}
,\ll{ilsg}\ee
 where \be g^2 (u)={1 \ov 2\sin^2 \al} \left ( {\kappa  }+ 
\sin   \al (s-u) (M^+ \sin   \al (u+s+1)
+{M^- } \cos \al (u+s+1 ) ) \right ) \ll{isg}\ee
 with  arbitrary constant $\kappa$, spin constant $s $ and central elements 
$ M^\pm=\pm   \sqrt {\pm 1} ( c^+_1c^-_2 \pm
c^-_1c^+_2 ) .$  
The Lax operator
(\re{Lt}) in realization (\re{ilsg}) of  its
underlying algebra  (\re{qa})
 would represent a generalized
 discrete sine-Gordon  model, which for the trivial choice
 of  $c^\pm_a={m \ov \sqrt 2}$, resulting $M^+=m^2, M^-=0, $  
 would recover the known exact lattice SG proposed by Korepin at al \c{izko}.
We introduce  now the inhomogeneities  by considering the 
central elements to be different
at different lattice points:
 $c^\pm_1= m_je^{i\al \th_j}, 
 \ \ c^\pm_2= m_je^{-i\al \th_j},$ which yields  
 $   M_j^+=m_j^2 \ M_j^-=0,$
with variable mass $m_j$ and constructs  a novel {\it inhomogeneous 
integrable lattice SG model}
linked  to the same   trigonometric 
 $R$ matrix \c{kulskly}.
It is note worthy that, in this inhomogeneous case    
 we find the expression for the quantum
determinant as $qdet L(\xi)=tr (L_n(\xi) \sigma^2 L^T_n(\xi
q^{-1})\sigma^2)=
1+S_j(\xi  q^{-1}+ \xi ^{-2} q),$
where $ S_j=( {1 \ov 4}m_j \De )^2$ with explicit site-dependent
inhomogeneity, 
which generalizes its  homogeneous case.
This leads to the nonlocal interactions of the Hamiltonian in both the
classical as well as quantum cases. Recall that in the corresponding 
homogeneous limit one gets nonlocality only in the quantum case \c{izko}.
Nevertheless, due to the property of the Lax-operator 
(\re{Lt}): $ \si ^2 L^*(\xi^*)\si ^2=L(\xi)$ the Hamiltonian of the
inhomogeneous lattice SG model as in the homogeneous case
 is also hermitian.
   
  For finding  the corresponding inhomogeneous field model 
 we may scale $p$ and $m_{j}$ 
by  lattice constant $\Delta$
and take the limit $\De \to 0.$ This would   derive from 
(\ref{Lt}) through $\sigma^1L_t
\to
I+\De {\cal L}$, the Lax operator $ {\cal L}$ of the   
 SG field  model 
with variable mass parameter $m=m(x,t)$. 
The corresponding  Hamiltonian may be expressed as 
\be{\cal H}= \int dx \left [ m(x,t) (u_t)^2 + (1/m(x,t)) (u_x)^2 +
 8(m_0-m(x,t)
\cos (2 \al u )) \right], \ll{msg1}\ee 
describing  an integrable inhomogeneous sine-Gordon field model with
variable mass and placed    in an external gauge field $\th (x)$.
 Such 
variable mass sine-Gordon equations may arise  in physical situations
\c{msg} and therefore the related exact results become  important.

\noindent {\bf IV. Quantum inhomogeneous Ablowitz-Ladik model}

 Considering in (\re{Lt}) a further  $p$-deformation of the spin operators:
 ${S}^\pm \to {\tilde S}^\pm = p^{- S^3} {S}^\pm $,
which corresponds  to a $p,q$ deformed trigonometric $R$-matrix,
and supposing $ c^-_1=c^+_2=0 $ and tuning $p=q$ , we get from the 
 second commutator of (\re{qa}) the  relation  
 \be 
(q^{-1} {\tilde S}^+ {\tilde S}^-- q {\tilde S}^-{\tilde S}^+)
= -h c^+_1c^-_2.
  \ll{qalm}\ee
If we choose  $c^+_1c^-_2=1$ and denote 
$ {\tilde S}^+=b, \  {\tilde S}^-=-b^\dag,$
(\ref{qalm}) reduces to a typical  q-oscillator algebra
 \c{kulskly}: $bb^\dag -q^2b^\dag b=q^2-1 $, which curiously yields the 
known commutator in the AL model:
$[b,b^\dag]= \hbar(1+ b^\dag b) $, by setting $\hbar=q^{2}-1 $.
Note that it is   easy to  introduce inhomogeneity into the quantum AL
model by  considering   
 time dependence of central elements as  $c^+_1={1\over c^-_2}=
e^{i\Gamma(t)}$, while keeping the required condition $c^+_1c^-_2=1$ unchanged.
This  changes   
the   Lax operator of the quantum AL model to its    inhomogeneous
version
 \be
L^{AL(inh)}_j{(\la)} = \left( \begin{array}{c}
 \Lambda  \ \ \quad 
   -b_j^\dag   \\
   b_j  \quad  
    \quad \ \ {1 \over \Lambda } 
          \end{array}   \right), \ \ \ \Lambda=\xi e^{i\Gamma(t)}, \ll{Lialm} \ee
which at the classical limit  
 coincides  obviously
 with  the Lax operator of \c{br3,bi2}.
Mutually commuting conserved operators $C_{\pm j}, j=N, N-1, \ldots , 1 $
 of this 
 inhomogeneous quantum AL model  can be obtained  
from the expansion of the transfer matrix $\tau(\la)$, constructed
 using the Lax operator (\re{Lialm}), while the  Hamiltonian is constructed from the 
operators $ C_{N-2}+C_{2-N}$ together with a term related to
 the quantum determinant of (\re{Lialm})
in the form  
\be H= \sum_n \tilde b^\dag_{n-1}\tilde b_n + 
 \tilde b^\dag_{n}\tilde b_{n-1} +{2\hbar \ov
 \log (1+\hbar) }\ \log (1+
\tilde b^\dag_{n}\tilde b_{n}), \ll{hal}
\ee
Here we have performed a canonical transformation $
 b_n \to \tilde b_n=be^{-i 2n\Gamma(t)} $ to remove explicit time dependence
from the Hamiltonian. Though Hamiltonian (\re{hal}) looks like 
the homogeneous one \c{kulskly}, 
due to  $\dot b_n \to \dot{\tilde b}_n+
{i 2n \dot \Gamma (t)}\tilde b_n $,
 an extra inhomogeneous term would appear in the
evolution equation and  therefore  using the
q-oscillator type commutation relation of AL model one can derive
from  Hamiltonian (\re{hal})     
 the known  inhomogeneous discrete NLS equation at the classical limit.


\noindent {\large {\bf 5. Exact Bethe ansatz solution}}

Exact solutions of the eigenvalue problem  for quantum integrable systems:
$\tau(\la)|m>=\Lambda_m(\la)|m> $, 
given through  the algebraic Bethe ansatz (ABA) \c{fadrev}, 
can be formulated almost in a model-independent way.  
The expression for the eigenvalues   may be 
given in the form 
\be \Lambda_m (\la) =
\al(\la) \pr_{j=1}^m {f(\la_j-\la)}  
+\bet(\la)
\pr_{j=1}^m {f(\la-\la_j)} 
, \ \mbox {where} \ {f(\la) }= {a(\la) \ov b(\la)}.
\ll{lambda}\ee
The Bethe $m$-particle eigenstates are defined as $|m>= \pr_{j=1}^m
B(\la_j)|0>$, with the pseudovacuum state satisfying $C(\la)|0>=0$, where
$B(\la) =T_{12} (\la), \ C(\la) =T_{21} (\la)$ are the off-diagonal elements 
of the monodromy matrix  $T(\la)=\prod_nL_n(\la)$.
 For the discrete (and also for the exact
 discretized versions of)
  quantum integrable models the Bethe momenta $\la_j$ are not arbitrary but
should be determined from the Bethe equations 
\be 
{\al(\la_j) \ov \bet (\la_j)}= \pr_{k \neq j}
 {a(\la_j-\la_k)\ov a(\la_k-\la_j) }  .\ll{be}\ee
Note that in  (\re{lambda}, \re{be})
 the coefficients $\al (\la)$ and $\bet (\la) $ are the pseudovacuum 
eigenvalues 
of the diagonal elements of $T(\la)$, i.e. $T_{11} (\la) \mid 0>= \al (\la)
\mid 0>, \ \ T_{22} (\la) \mid 0>= \bet (\la)
\mid 0>$ and therefore are  the only  model-dependent elements, since
they are related to the Lax operator $L_j(\la)$ of a concrete model.
On the other hand  the factors containing the function $ {f(\la-\la_j)} $
, which are  the major contributors in the above expressions, are given by
the ratio of the $R$-matrix elements:
$a(\la)= R^{11}_{11}(\la)$ and $ b(\la)= R^{12}_{12}(\la)$ and therefore
  depend not on individual models, but on the class to which the models
belong. Consequently, they remain 
  the same for all  models of the same class e.g., rational, trigonometric
etc.

Therefore, we can solve the eigenvalue problem of
 the quantum inhomogeneous   models by using the same formulas
(\re{lambda}, \re{be}) and even taking the factors $ {f(\la)} $  same as 
   their corresponding homogeneous counterparts, since in our construction 
we could keep the  quantum $R$-matrix same for both these cases.
 We have to
remember however that, since  the Lax operators are  changed
with the inclusion of inhomogeneity parameters, the expressions for 
 $\al (\la) , \bet (\la)$ would be more complicated. Therefore for the  
Bethe ansatz solution of  the exact lattice version
of  our inhomogeneous quantum NLS 
   we may use  (\re{lambda}, \re{be}), taking  the high spin limit
of $ \al (\la)=\prod_n (g_n(\la+s_n) +f_n),
\bet(\la)=\prod_n (g_n(\la-s_n) +f_n) $ , which follows from the
related Lax operator (\re {Linls}).

 The model independent part
for the NLS model as well as for the TC model, linked  to 
 the rational $R$-matrix,  should naturally 
be given by $a(\la)= \la +\alpha , \  b(\la)=
\la $.   The model-dependent part  for the quantum
inhomogeneous  TC model on the other hand, 
as evident from its Lax operator (\re {qitc}), is  to be taken as
  $ \al (\la)=\prod_n (\la +g_{2(n)}),
\bet(\la)=0 $. However  the basic problem associated with the TC model,
namely  the non-availability of the pseudovacuum, 
 remains   the same here as in    the homogeneous case.
  Therefore for the  solution of  the eigenvalue problem in the
 inhomogeneous  quantum TC model   one has to adopt also 
 the functional Bethe ansatz
method \c{fba}, in place of the above  algebraic approach.

For   inhomogeneous quantum lattice SG and  AL models also 
   we can use  the above 
  Bethe ansatz result and  since both of these   models are associated
  with  the   trigonometric $R_{trig}$-matrix (\re{Rtrig}),
 we have  $a(\la)= \sin(\la +\alpha), \ 
\ b(\la)=\sin(\la) $.
For inhomogeneous SG one has to use the pseudovacuum by taking the
product of two adjacent sites, as in its homogeneous case \c{fad}, though the
relevant details would be different here, since the mass $m_j$ becomes
site-dependent.

For the inhomogeneous AL model due to an additional
$p=q$-deformation one should use  $ 
\ b_\pm(\la)=q^\pm b(\la)$   in the above formulas
and   extra factors 
of $q^{-1}$ and $q$ should appear
 in the first and the second  term of 
(\re{lambda}),  respectively. Consequently 
  another 
 factor of  $q^{2}$ also  appears in the r.h.s. of
the Bethe equations (\re{be}). The rest of the terms 
needed for  the solution of   this inhomogeneous AL  model
 can be derived from its quantum Lax operator (\re{Lialm}) 
giving
  $ \al (\la)= (\xi e^{i\Gamma(t)})^N, \
\bet(\la)= (\xi e^{i\Gamma(t)})^{-N} $.

 \noindent {\large {\bf 6. Concluding remarks}}
 
We have proposed  quantum integrable extensions of  the inhomogeneous
NLS model, Toda chain and the Ablowitz-Ladik model, different 
forms of which are well known
  only at the classical level. 
We have also proposed an inhomogeneous SG model with variable mass,
 both at the exact lattice
and the field limit, which are novel models at classical as well as at the 
 quantum
level.
We have constructed 
such inhomogeneous quantum models 
 exploiting the Yang-Baxter equation and 
avoiding the non-isospectrality problem by using central elements of 
the underlying algebra
for introducing the inhomogeneity  in a unified way. We have also   
indicated how to get the exact
eigenvalue solutions of such quantum integrable inhomogeneous models systematically   
 through  the Bethe ansatz.
It has been found that the classical limits of the quantum inhomogeneous
models constructed here are close to   known inhomogeneous equations,
 though their  conventional forms can be  reached only after some 
noncanonical transformations, 
 permitted only at the classical level.
The method presented here should be applicable for constructing 
inhomogeneous extension of some other 
quantum integrable models like the Liouville model,
relativistic Toda chain etc. 

\noindent  {\bf Acknowledgment}

The author  likes to thank Prof Orlando Ragnisco of  Univ. Rome-tre for many
fruitful discussions.


 \end{document}